\documentclass[twocolumn,amsmath,amssymb,prb]{revtex4}
\usepackage{graphicx}% Include figure files
\usepackage{dcolumn}% Align table columns on decimal point
\usepackage{bm}% bold math
\usepackage{color}

\def \be {\begin{equation}}
\def \ee {\end{equation}}
\def \ben {\begin{eqnarray}}
\def \een {\end{eqnarray}}

\begin{document}

\bibliographystyle{../prsty}

\title{Real time quantum dynamics preaveraged over imaginary time path integral: A formal basis for both Centroid Molecular Dynamics and Ring Polymer Molecular Dynamics }

\author{Seogjoo Jang}
\affiliation{Department of Chemistry and Biochemistry, Queens College and the Graduate Center, City University of New York, 65-30 Kissena Boulevard, Flushing, New York 11367  }

\date{\today}

\begin{abstract}
An exact real time quantum dynamics preaveraged over imaginary time path integral is formulated for general condensed phase equilibrium ensemble.  This formulation results in the well-known centroid dynamics approach upon filtering of centroid constraint, and provides a rigorous framework to understand and analyze a related quantum dynamics approximation method called ring polymer molecular dynamics. The formulation also serves as the basis for developing new kinds of quantum dynamics that utilize the cyclic nature of  the imaginary time path integral. 

\end{abstract}

%\pacs{82.20.Xr, 05.60.Gg, 82.20.Ln, 03.65.Sq}

\maketitle
While Feynman path integral\cite{feynman-qmech,kleinert-pi4} paved the way to routine numerical simulation of equilibrium static quantum properties of condensed phase systems,\cite{berne-anp37,doll-acp77,schmidt-book1} its application to the calculation of real time propagator has been plagued by so called sign problem.  Although ingenious numerical techniques\cite{doll-acp73,moreira-prl91,makri-arpc50,mak-jcp131} to ameliorate the resulting convergence issue have been developed, a general method applicable to condensed phase molecular systems with good scalability is absent at present.  On the other hand, major equilibrium dynamical  properties in condensed media can be expressed in terms of time correlation functions averaged over equilibrium density operator, for which the sign problem is less severe.  Various approaches and approximation methods to calculate quantum time correlation functions\cite{gallicchio-jcp105,kim-jcp108,cao-jcp99,voth-acp93,poulsen-jcp119,craig-jcp121,craig-jcp122,georgescu-prb82,liu-jcp134-1} have been developed,  but accurate calculation of coherent features beyond thermal time scale for general molecular system remains a very challenging problem. 
Considering recent experimental evidences\cite{engel-nature466,collini-nature463} that long lasting quantum coherence can be significant even in complex biological systems, reliable calculation of real time quantum correlation functions for sufficiently long time can be crucial for elucidating important quantum dynamical effects hidden in complex condensed phase systems.  Path integral approach has unique advantage in such effort because it allows trajectory based all-atomistic simulation that scales favorably with the size of the system. The present communication constructs a general formalism that utilizes this merit to its maximum extent through preaveraging of time correlation function over imaginary time path integral.  This provides a common theoretical basis for both Centroid Molecular Dynamics (CMD)\cite{cao-jcp99,jang-cen1,jang-cen2} and ring polymer molecular dynamics (RPMD),\cite{craig-jcp121,craig-jcp122}  and offers new theoretical insights for improving such methods.

Consider the  standard Hamiltonian given by $\hat H=\hat p^2/(2m)+V(\hat q)$ assuming one dimension for notational convenience.  
Factoring the canonical density operator $e^{-\beta \hat H}$ into $N$ pieces of $e^{-\beta \hat H/N}$ and making the approximation of symmetric Trotter factorization for each factor, $e^{-\beta \hat H} \approx \left( e^{-\beta V(\hat q)/(2N)} e^{-\beta \hat T/N } e^{-\beta V(\hat q)/(2N)}\right)^N$.
Representing $e^{-\beta V(\hat q)/(2N)}$ in the position basis of $\hat 1=\int dq_k |q_k\rangle\langle q_k|$, and $e^{-\beta \hat T/N}$ in the momentum basis of $\hat 1=\int dp_k |p_k\rangle\langle p_k|$, we obtain the following standard phase space path integral expression\cite{kleinert-pi4} for the canonical density operator:\footnote{In the present work, any approximate identity that becomes exact in the limit of $N\rightarrow \infty $ will be represented by the equality sign.}  
\ben
&&e^{ - \beta \hat H}= \left (\frac{1}{2\pi\hbar}\right)^{N}\int \cdots \int dq_1\cdots dq_{_{N+1}} dp_1\cdots dp_{_N}  \nonumber \\
&&\mbox{\makebox[0.2 in]{ }}|q_1\rangle \prod_{k=1}^{N}\left\{e^{-\beta V( q_k)/(2N)} e^{-\beta p_k^2/(2mN)} e^{-\beta V(q_{k+1})/(2N)} \right . \nonumber \\
&&\mbox{\makebox[1. in]{ }}\times \left . e^{ip_k(q_k-q_{k+1})/\hbar} \right\}\langle q_{_{N+1}}|\ . \label{eq:can_path}
\een
With the replacement of  $(q_1+q_{_{N+1}})/2 \rightarrow q_1$ and $q_1-q_{_{N+1}}\rightarrow \eta$, Eq. (\ref{eq:can_path}) can be recast into the following form: 
\ben
&&e^{-\beta\hat H}= \int  d\eta \int d{\bf q} \int d{\bf p}\  U({\bf q},{\bf p})\nonumber \\
&&\hspace{.5in} \times J(q_1,\frac{p_1+p_{_N}}{2};\eta) |q_1+\frac{\eta}{2}\rangle  \langle q_1-\frac{\eta}{2}|\ , \label{eq:can_n0}
\een
where ${\bf q}\equiv (q_1,\cdots,q_{_N})$, ${\bf p}\equiv(p_1,\cdots,p_{_N})$, 
\ben
&&U({\bf q}, {\bf p})=\left(\frac{1}{2\pi\hbar}\right)^N \prod_{k=1}^{N}\left\{e^{-\beta V( q_k)/N}\right .\nonumber \\ 
&&\mbox{\makebox[.5 in]{ }}\left . \times e^{-\beta p_k^2/(2mN)} e^{ip_k(q_k-q_{k+1})/\hbar} \right\}   \label{eq:u_def}\ , \\
&&J(q,\frac{p_1+p_{_N}}{2};\eta)= e^{-\beta \Delta V(q;\eta)/N} e^{\frac{i\eta}{2\hbar} (p_1+p_{_N})}  \label{eq:j_def}\ .
\een
In Eq. (\ref{eq:u_def}), $q_{_{N+1}}=q_1$.  
In Eq. (\ref{eq:j_def}), $\Delta V (q;\eta)=(V(q+\eta/2)+V(q-\eta/2))/2-V(q)$.

The path integration in Eq. (\ref{eq:can_n0}) is independent of the cyclic permutation of the dummy variables,  
$q_k \rightarrow q_{k+l}$ and $p_k \rightarrow p_{k+l}$, under the cyclic boundary condition of $q_{_{N+k}}=q_k$ and $p_{_{N+k}}=p_k$.  The integrand $U({\bf q},{\bf p})$ defined by Eq. (\ref{eq:u_def}) is also invariant with this permutation.   Thus, Eq. (\ref{eq:can_n0}) can be  symmetrized as follows: 
\be
e^{-\beta\hat H}= \int d{\bf q} \int d{\bf p}\ U({\bf q},{\bf p}) \hat S({\bf q}, {\bf p}) \ , \label{eq:e_beta_s}
\ee
where 
\be
\hat S({\bf q},{\bf p})=\frac{1}{N}\sum_{k=1}^N \int d \eta \ J(q_k,\frac{p_k+p_{k-1}}{2};\eta)|q_k+\frac{\eta}{2}\rangle\langle q_k-\frac{\eta}{2}| \ .\label{eq:s_def}
\ee
In the above equation, it is understood that $p_{_0}=p_{_N}$.  

The operator defined by Eq. (\ref{eq:s_def}) has unit trace and satisfies the following identities:
\ben
&&Tr\left\{\hat S({\bf q},{\bf p}) \hat A\right\}=\frac{1}{N}\sum_{k=1}^N A(q_k)\equiv A_0({\bf q}) \ , \label{eq:a0_def} \\
&&Tr\left\{\hat S({\bf q},{\bf p})\hat p\right\}=\frac{1}{N}\sum_{k=1}^N p_k \equiv p_0({\bf p})\  ,
\een
where $\hat A=A(\hat q)$ is an operator depending on position only, and $A_0({\bf q})$ and $p_0({\bf p})$ are centroids of $A_k$'s and $p_k$'s respectively.
Thermal averages of $A(\hat q)$ and $\hat p$ can be expressed as the averages of these centroids  over the complex distribution function $U({\bf q},{\bf p})$ defined by Eq. (\ref{eq:u_def})  as follows:  
\ben 
&&Tr\left\{ e^{-\beta \hat H} A(\hat q)\right\}= \int d{\bf q} \int d{\bf p}\ U({\bf q},{\bf p}) A_0({\bf q}) \ , \\
&&Tr\left\{ e^{-\beta \hat H} \hat p\right\}= \int d{\bf q} \int d{\bf p}\ U({\bf q},{\bf p})p_0({\bf p})\ .
\een

Dynamical extension of the above expressions is possible for Kubo-transformed time correlation functions, which are generic forms for quantum transport coefficients within the linear response theory.\cite{kubo-jpsj-12} 
Consider the following correlation between $\hat A$ and an arbitrary operator $\hat B$: 
\be
\tilde C_{AB}(t)=\frac{1}{Z}\int_0^\beta \frac{d\lambda}{\beta} Tr\left\{ e^{-\lambda \hat H}\hat A e^{-(\beta-\lambda) \hat H}e^{i\hat H t/\hbar} \hat B e^{-i\hat H t/\hbar}\right\} \ , \label{eq:c_abt0}
\ee
where $Z=Tr\{e^{-\beta \hat H}\}$.  The integration over $\lambda$ in Eq. (\ref{eq:c_abt0}) can be approximated as
\ben
&&\int_0^\beta \frac{d\lambda}{\beta} e^{-\lambda \hat H}\hat A e^{-(\beta-\lambda) \hat H}= \frac{1}{2N}\left (\hat A e^{-\beta \hat H}+e^{-\beta\hat H}\hat A\right) \nonumber \\
&&\mbox{\makebox[.5 in]{ }}+\frac{1}{N}\sum_{j=1}^{N-1} e^{-\beta(j/N)\hat H}\hat A e^{-\beta (1-j/N)\hat H} \ .\label{eq:kubo_n} 
\een
Approximating each $e^{-\beta \hat H/N}$ in the above expression by the 
symmetric Trotter factorization, $e^{-\beta \hat T/(2N)}e^{-\beta \hat V/N}e^{-\beta \hat T/(2N)}$, and going through the same procedure that leads to Eq. (\ref{eq:e_beta_s}) from Eq. (\ref{eq:can_n0}), we obtain
\ben
&&\frac{1}{Z}\int_0^\beta \frac{d\lambda}{\beta} e^{-\lambda \hat H} \hat A e^{-(\beta-\lambda)\hat H}\nonumber 
\\ && = \frac{1}{Z}\int d{\bf q}\int d{\bf p}\ U({\bf q},{\bf p})\hat S ({\bf q},{\bf p})A_0({\bf q}) \ , \label{eq:a0_kubo}
\een
where $\hat S ({\bf q},{\bf p})$ and $A_0({\bf q})$ are defined by Eqs. (\ref{eq:s_def}) and (\ref{eq:a0_def}).
Inserting Eq. (\ref{eq:a0_kubo}) into Eq. (\ref{eq:c_abt0}), we obtain
\be
\tilde C_{AB}(t)= \frac{1}{Z}\int d{\bf q}\int d{\bf p}\ U({\bf q},{\bf p})A_0 ({\bf q}) Tr \left\{ \hat S(t;{\bf q},{\bf p}) \hat B \right\} \label{eq:cabt-1} \ , 
\ee
where
\be
\hat S (t;{\bf q},{\bf p})\equiv e^{-i\hat Ht/\hbar}\hat S({\bf q},{\bf p})e^{i\hat H t/\hbar} \ . \label{eq:st}
\ee
This is a time-dependent version of Eq. (\ref{eq:s_def}) and is governed by the quantum Liouville equation like any normal density operator.  Thus, in terms of Eq. (\ref{eq:st}), we can define the following time dependent ``average" of $\hat B$: 
\be
B_0(t;{\bf q},{\bf p}) =Tr\{\hat S(t;{\bf q}, {\bf p}) \hat B\}\ .  \label{eq:bqt_def}
\ee
Then, Eq. (\ref{eq:cabt-1}) reduces to a form that involves a classical-like correlation between $A_0({\bf q})$ and $B_0(t;{\bf q},{\bf p})$.

Now suppose we can identify an appropriate vector function ${\bf x}_0({\bf q},{\bf p})$, which depends on the phase space path $({\bf q}, {\bf p})$,  and a relevant filtering function,\cite{doll-acp73} $F ({\bf x}-{\bf x}_0({\bf q},{\bf p}))$,  which is normalized as follows: 
\be
1=\int d{\bf x}\ F( {\bf x}-{\bf x}_0({\bf q},{\bf p}))\ .
\ee  
Inserting the above identity into Eq. (\ref{eq:cabt-1}) and defining 
\be 
\rho_0({\bf x})=\int d{\bf q} \int d{\bf p}\ F( {\bf x}-{\bf x}_0({\bf q},{\bf p})) U({\bf q},{\bf p})\ , 
\ee
we obtain the following expression: 
\be
\tilde C_{AB}(t)= \frac{1}{Z}\int d{\bf x}\ \rho_0({\bf x}) \langle A_0 ({\bf q}) B_0 (t;{\bf q},{\bf p}) \rangle_{\bf x} \ ,\label{eq:c_abt-1}
\ee
where the definition of Eq. (\ref{eq:bqt_def}) has been used and 
\be 
\langle ... \rangle_{\bf x}\equiv\int d{\bf q}\int d{\bf p} \frac{F ({\bf x}- {\bf x}_0({\bf q},{\bf p}))}{\rho_0 ({\bf x})} \ U({\bf q},{\bf p}) (... )\ .
\ee
It is easy to show that $Z=\int d{\bf x}\ \rho_0({\bf x})$.  Thus, $\rho_0({\bf x})/Z$ can serve as a genuine probability density given that the filtering function $F({\bf x}-{\bf x}_0({\bf q},{\bf p}))$  can be chosen to make $\rho_0({\bf x})$ positive definite. 
For $\hat p$, a similar expression can be obtained as follows. 
\be
\tilde C_{pB}(t)= \frac{1}{Z} \int d{\bf x}\ \rho_0({\bf x}) \langle p_0 ({\bf p}) B_0 (t;{\bf q},{\bf p}) \rangle_{\bf x} \label{eq:c_pbt} \ . 
\ee

Equations (\ref{eq:c_abt-1}) and (\ref{eq:c_pbt}) are exact except for the discretization error of the path integral, which vanishes in the limit of $N\rightarrow \infty$.  These serve as  general expressions that utilize the cyclic nature of the imaginary paths, and can be used as common framework to understand the CMD\cite{cao-jcp99,jang-cen1,jang-cen2} and the RPMD\cite{craig-jcp122} methods as detailed below. 

Let us consider the case where ${\bf x}_0({\bf q},{\bf p})=(q_0({\bf q}),p_0({\bf p}))$ and 
\be
F({\bf x}_c-{\bf x}_0({\bf q},{\bf p}))\equiv \delta (q_c-q_0({\bf q}))\delta (p_c-p_0({\bf p})) \ .\label{eq:filter-centroid}
\ee  
Then, $\rho_0({\bf x})=\rho_c(q_c,p_c)$, which is the phase space centroid density.\cite{jang-cen1}   For the case where  $\hat A=\hat q$, Eq. (\ref{eq:c_abt-1}) becomes
\be
\tilde C_{qB}(t)=\frac{1}{Z} \int \int dq_c dp_c\ \rho_c(q_c,p_c) q_c B_c(t;{\bf q},{\bf p})\ ,
\ee 
where 
\ben
B_c(t;{\bf q}, {\bf p}) &=&\int d{\bf q} \int d{\bf p} \frac{\delta (q_c-q_0)\delta (p_c-p_0)}{\rho_c(q_c,p_c)}\nonumber \\
&&\hspace{.5in} \times U({\bf q},{\bf p}) B_0(t;{\bf q},{\bf p}) \nonumber \\ 
&=&Tr\{\hat \delta_c(t;q_c,p_c) \hat B\} \ .
\een
In the above equation, $\hat \delta_c(q_c,p_c;t)$ is the time dependent quasi-density operator defined in the formulation of centroid dynamics.\cite{jang-cen1}  In the present notation,  it is expressed as 
\ben 
\hat \delta_c(t;q_c,p_c)& \equiv& \int d{\bf q} \int d{\bf p} \frac{\delta (q_c-q_0)\delta (p_c-p_0)}{\rho_c(q_c,p_c)} \nonumber \\
&&\hspace{.5in}\times U({\bf q},{\bf p})\hat S(t;{\bf q},{\bf p})\ .
\een
Similarly, it is straightforward to show that Eq. (\ref{eq:c_pbt}) can be expressed as 
\be 
\tilde C_{pB}(t)=\frac{1}{Z} \int \int dq_c dp_c\ \rho_c(q_c,p_c) p_c B_c(t;{\bf q},{\bf p}) \ .
\ee
Thus, for the filtering function of Eq. (\ref{eq:filter-centroid}), Eqs. (\ref{eq:c_abt-1}) and (\ref{eq:c_pbt}) become equivalent to the exact centroid dynamics,\cite{jang-cen1} which is well defined when $\hat A$ is a linear combination of $\hat q$ and $\hat p$.  The CMD method\cite{cao-jcp99} amounts to further semiclassical approximation\cite{jang-cen2} for the time evolution of $B_c(t;{\bf q}, {\bf p})$.  

As another choice, consider the case where ${\bf x}_0({\bf q},{\bf p})={\bf q}$ and 
\be
F({\bf x}-{\bf x}_0 ({\bf q},{\bf p}))\equiv \delta ({\bf x}-{\bf q})\ .
\ee  
With this choice, $\rho_0({\bf x})=\int d{\bf  p}\ U({\bf x},{\bf p})\equiv R({\bf x})$, which has the form of a ring polymer position-distribution and is given by 
\be
R({\bf x})=\left (\frac{mN}{2\pi\beta\hbar^2}\right)^{N/2} \prod_{k=1}^N e^{-\frac{\beta}{N} V (x_k)} e^{-\frac{m N}{2\beta\hbar^2} (x_k-x_{k+1})^2}\ .
\ee
Then, Eq. (\ref{eq:c_abt-1}) reduces to 
\be
\tilde C_{AB}(t)= \frac{1}{Z}\int d{\bf x}\  R({\bf x}) A_0 ({\bf x}) \langle B_0(t;{\bf q},{\bf p})\rangle_{\bf x}  \ ,\label{eq:c_abt-2}
\ee
where
\ben
\langle B_0(t;{\bf q},{\bf p})\rangle_{\bf x}=\int d{\bf p}\ M_{\bf x}({\bf p}) Tr \left \{ \hat S (t;{\bf x}, {\bf p}) \hat B\right\} \ . \label{eq:bxt-1}
\een
In the above expression, 
\ben 
&&M_{\bf x}({\bf p})\equiv \frac{U({\bf x},{\bf p})}{R({\bf x})}\nonumber \\
&&=\left (\frac{\beta}{2\pi m N} \right)^{N/2} \prod_{k=1}^N e^{-\frac{\beta}{2mN} (p_k -i\frac{mN}{\beta\hbar}(x_k-x_{k+1}))^2}\ , \nonumber \\ 
\een
and
\ben
&&Tr\left \{\hat S (t;{\bf x},{\bf p}) \hat B \right\} =\frac{1}{N} \sum_{k=1}^N \int d\eta J(x_k,\frac{p_k+p_{k+1}}{2};\eta) \nonumber \\
&&\hspace{.7in} \times \langle x_k-\frac{\eta}{2}| e^{i\hat H t/\hbar} \hat B e^{-i\hat H t/\hbar} |x_k+\frac{\eta}{2}\rangle \ . \label{eq:sxpt_b}
\een

It is instructive to calculate all the terms in Eq. (\ref{eq:sxpt_b}) explicitly for a harmonic oscillator with Hamiltonian $\hat H_\omega=\hat p^2/(2m)+m\omega^2 \hat q^2/2$.  For this case, the $\Delta V(q;\eta)$ introduced below Eq. (\ref{eq:j_def}) becomes $m\omega^2\eta^2/8$, and  as a result, Eq. (\ref{eq:j_def}) reduces to
\be
J_\omega (x_k,\frac{p_k+p_{k+1}}{2};\eta)=e^{-\frac{\beta m\omega^2}{8N}\eta^2} e^{\frac{i\eta}{2\hbar}(p_k+p_{k+1})} \ , \label{eq:j_omg}
\ee
and the time dependent matrix element in Eq. (\ref{eq:sxpt_b}) can be expressed as
\ben
&&\langle x_k-\frac{\eta}{2}|e^{i\hat H_\omega t/\hbar}\hat B e^{-i\hat H_\omega t/\hbar} |x_{k}+\frac{\eta}{2}\rangle \nonumber \\
&&=\frac{m\omega}{2\pi \hbar|\sin (\omega t)|}\int dx'\int dx'' \langle x'|B|x''\rangle \nonumber  \\
&&\times \exp\Big\{\frac{im\omega}{2\hbar \sin (\omega t)} \big( (x''^2-x'^2) \cos (\omega t)+2\eta x_k \cos (\omega t) \nonumber \\
&&\hspace{.5in}  -2 (x''-x') x_k -(x''+x')\eta \big )\Big \} \ . \label{eq:bx_omg}
\een
Inserting Eqs. (\ref{eq:j_omg}) and (\ref{eq:bx_omg}) into Eq. (\ref{eq:sxpt_b}), introducing $X=(x'+x'')/2$ and $x=x'-x''$, and integrating the resulting expression over $\eta$, we can obtain the following expression: 
\ben 
&&Tr\left \{\hat S_\omega (t;{\bf x},{\bf p})\hat B \right\}\nonumber \\
&&=\frac{1}{N} \sum_{k=1}^N \int dX \int dx  \langle X+\frac{1}{2} x|\hat B|X-\frac{1}{2}x\rangle \nonumber \\
&&\times  \left (\frac{2 mN}{\pi \beta \hbar^2 \sin^2 (\omega t)}\right)^{1/2} e^{-\frac{im\omega}{\hbar \sin (\omega t)}x  (X\cos(\omega t)-x_k)}\nonumber \\
&&\times e^{-\frac{2mN}{\beta \hbar^2 \sin^2 (\omega t)} (X-x_k \cos (\omega t)-\frac{p_k+p_{k+1}}{2m\omega} \sin (\omega t)  )^2  }\ . \label{eq:sbt_omg}
\een
For the case where $\hat B$ is an operator depending only on position, Eq. (\ref{eq:sbt_omg}) simplifies to 
\ben 
&&Tr\left \{\hat S_\omega (t;{\bf x},{\bf p})B (\hat x) \right\}\nonumber \\
&&=\frac{1}{N} \sum_{k=1}^N \int dX  B(X) \left (\frac{2 mN}{\pi \beta \hbar^2 \sin^2 (\omega t)}\right)^{1/2} \nonumber \\
&&\times e^{-\frac{2mN}{\beta \hbar^2 \sin^2 (\omega t)} (X-x_k \cos (\omega t)-\frac{p_k+p_{k+1}}{2m\omega} \sin (\omega t)  )^2  }\ .  \label{eq:sbxt}
\een  
The above expression represents a swarm of time dependent Gaussian functions with their centers evolving according to the classical equation of motion.  
This is similar to the prescription of the RPMD method but is not the same as will be noted below.

Consider the case where $\hat B=\hat q$.  Then, the integration in Eq. (\ref{eq:sbxt}) simply results in the classical trajectories of harmonic oscillator, and the time correlation function defined by Eq. (\ref{eq:c_abt-2}) becomes 
\ben
\tilde C_{Aq}^\omega(t)= \frac{1}{Z} \int d{\bf x} \int d {\bf p}\ R({\bf x})M_{\bf x}({\bf p})A_0({\bf x})\nonumber \\
\times \frac{1}{N} \sum_{k=1}^N (x_k\cos (\omega t)+\frac{p_k+p_{k+1}}{2m\omega} \sin (\omega t) )\ .
\een    
Assuming  that $p_k$ can be a complex value and replacing it with $\tilde p_k+i\frac{mN}{\beta\hbar}(x_k-x_{k+1})$, where $\tilde p_k$ is real, we obtain 
 \be
\tilde C_{Aq}^\omega(t)= \frac{1}{Z} \int d{\bf x} \int d \tilde {\bf p}\ R({\bf x})M_{0}(\tilde {\bf p})A_0({\bf x}) x_0(t) \ , \label{eq:caq_omega_t}
\ee
where
\be
x_0(t)=\frac{1}{N} \sum_{k=1}^N (x_k\cos (\omega t)+\frac{\tilde  p_k+\tilde p_{k+1}}{2m\omega} \sin (\omega t) )\ . 
\ee    
In the above expression, $M_0(\tilde {\bf p})$ is the classical Maxwell-Boltzmann distribution for $\tilde {\bf p}$ and the fact that $\sum_{k=1}^N(x_k-x_{k+1})=0$ has been used.  

Equation (\ref{eq:caq_omega_t}) is equivalent to the prescription of the RPMD method except for the slight difference that $x_0(t)$ evolves from the initial momentum $(\tilde {p}_k+\tilde p_{k+1})/2$.   Although the RPMD method includes additional unphysical force term due to the ring polymer harmonic potential, $\sum_k \frac{mN}{2\beta^2\hbar^2} (x_k-x_{k+1})^2$,  their net effect on the time evolution of the centroid $x_0(t)$ is zero.   Thus, Eq. (\ref{eq:caq_omega_t}) can be considered as the direct quantum mechanical derivation of the RPMD method for harmonic oscillator when $\hat B$ is linear in position. The case when $\hat B$ is linear in momentum can also be obtained taking time derivative of this expression.  It is important to note that the derivation shown above is distinct from the original\cite{craig-jcp121} or follow-up\cite{braams-jcp125} justification of the RPMD method because any of these does not start from the consideration of quantum mechanical density operator and is limited in its scope.  The present analysis shows that if the physical observable  $\hat B$ is nonlinear function of position, the ring polymer force term makes nontrivial contribution to the time evolution of the centroid $B_0$ even for harmonic oscillator.   Thus, this clarifies why the RPMD method is not exact even for harmonic oscillators if $\hat B$ is a nonlinear function of position.    

For general Hamiltonian and operator $\hat B$, implementing similar translation of the momenta in the integrand Eq. (\ref{eq:bxt-1}), we can express Eq. (\ref{eq:c_abt-2}) as follows:
\ben
&&\tilde C_{AB}(t)=\frac{1}{Z} \int d{\bf x} \int d\tilde {\bf p} R({\bf x}) M_0(\tilde {\bf p}) \nonumber \\
&&\hspace{.5in} \times  A_0({\bf x}) Tr\{ \hat S(t;{\bf x},\tilde {\bf p}+i{\bf p}_i)\hat B\}\ , \label{eq:cabt-ex}
\een 
where ${\bf p}_i$ is the imaginary term of the complex momentum ${\bf p}$ with the following component: $({\bf p}_i)_{k}=\frac{mN}{\beta\hbar} (x_k-x_{k+1})$.  Then, formally, the RPMD method amounts to the following approximation: 
\be
Tr\{ \hat S(t;{\bf x},\tilde {\bf p}+i{\bf p}_i)\hat B\}\approx  Tr\{ \hat S({\bf x}(t),\tilde {\bf p}(t))\hat B\} \ ,
\ee
\vspace{.1in}\\ 
where ${\bf x}(t)$ and $\tilde {\bf p}(t)$ satisfy the RPMD equations of motion.\cite{craig-jcp121}    This prescription corresponds to the exact time evolution for the position of the harmonic oscillator as demonstrated by Eq. (\ref{eq:caq_omega_t}), and guarantees producing correct equilibrium ensemble and classical limit.  However, as noted above, the introduction of force due to to the ring polymer harmonic potential term $\sum_k \frac{mN}{2\beta^2\hbar^2} (x_k-x_{k+1})^2$, which does not have any quantum dynamical origin even for harmonic oscillator, can introduce artifacts for nonlinear operators and anharmonic systems.  Indeed, this issue has already been observed numerically in a work reporting spurious vibrational peaks in the RPMD method.\cite{witt-jcp130}  The analysis given above elucidates the source of such observation at formal level. 

In summary, the formalism of the present communication provides a common theoretical framework for understanding the CMD and RPMD methods as means to calculate  Kubo-transformed time correlations functions.   In particular, the present formalism allows clear understanding of quantum dynamical approximation involved in the RPMD method, for which no rigorous quantum dynamical derivation has been provided so far.  Its exact counterpart, Eq. (\ref{eq:cabt-ex}), can be used to examine the accuracy of its assumptions against exact or well defined semiclassical calculation methods.\cite{miller-jpca105,makri-arpc50}  In addition, the general expressions, Eqs. (\ref{eq:c_abt-1}) and (\ref{eq:c_pbt}), offer new possibility to develop different kinds of approximation methods given that  new and effective filtering functions can be identified.   Finally, direct evaluation of Eq. (\ref{eq:sxpt_b}) opens up the possibility of combining semiclassical methods\cite{miller-jpca105,makri-arpc50} with the imaginary time path integral for the evaluation of equilibrium time correlation functions without going through the Wigner distribution.

\acknowledgments

This work was supported by the National Science Foundation CAREER award (Grant No. CHE-0846899), the Office of Basic Energy Sciences, Department of Energy (Grant No. DE-FG02-09ER16047), and the Camille Dreyfus Teacher Scholar Award.  The author acknowledges Anton Sinitskiy for identifying some typos in the first version of this manuscript.

\end{document}